\documentclass[aps,prb,showpacs,amsmath,amssymb,superscriptaddress,twocolumn]{revtex4-1}

\usepackage{graphicx}
\usepackage{dcolumn}
\usepackage{bm}
\usepackage{siunitx}

\begin{document}

\preprint{APS/123-QED}

\title{Josephson junctions in double nanowires bridged by in-situ deposited superconductors}


 \author{Alexandros Vekris}
 \affiliation{Center For Quantum Devices, Niels Bohr Institute, University of Copenhagen, 2100 Copenhagen, Denmark}
 \affiliation{Sino-Danish College (SDC), University of Chinese Academy of Sciences}
 
 \author{Juan Carlos Estrada Salda\~na} 
 \affiliation{Center For Quantum Devices, Niels Bohr Institute, University of Copenhagen, 2100 Copenhagen, Denmark}
  
 \author{Thomas Kanne}
 \affiliation{Center For Quantum Devices, Niels Bohr Institute, University of Copenhagen, 2100 Copenhagen, Denmark}
  
\author{Mikelis Marnauza} \author{Dags Olsteins}
\affiliation{Center For Quantum Devices, Niels Bohr Institute, University of Copenhagen, 2100 Copenhagen, Denmark}

\author{Furong Fan} 
\affiliation{Beijing Key Laboratory of Quantum Devices, Key Laboratory for the Physics and Chemistry of Nanodevices and Department of Electronics, Peking University, Beijing 100871, China}

\author{Xiaobo Li} 
\affiliation{Beijing Key Laboratory of Quantum Devices, Key Laboratory for the Physics and Chemistry of Nanodevices and Department of Electronics, Peking University, Beijing 100871, China}

\author{Thor Hvid-Olsen}
\affiliation{Center For Quantum Devices, Niels Bohr Institute, University of Copenhagen, 2100 Copenhagen, Denmark}

\author{Xiaohui Qiu}
\affiliation{CAS Key Laboratory
of Standardization and Measurement for Nanotechnology, National Center for Nanoscience and Technology, Beijing 100190, China}
\affiliation{CAS Center for Excellence in Nanoscience, National Center for Nanoscience and Technology, Beijing 100190, China}

\author{Hongqi Xu} 
\affiliation{Beijing Key Laboratory of Quantum Devices, Key Laboratory for the Physics and
Chemistry of Nanodevices and Department of Electronics, Peking University, Beijing
100871, China}
\affiliation{Beijing Academy of Quantum Information Sciences, 100193 Beijing, China}

\author{Jesper Nyg{\aa}rd}
\affiliation{Center For Quantum Devices, Niels Bohr Institute, University of Copenhagen, 2100 Copenhagen, Denmark}
    
\author{Kasper Grove-Rasmussen}
\email{k\_grove@nbi.ku.dk}
\affiliation{Center For Quantum Devices, Niels Bohr Institute, University of Copenhagen, 2100 Copenhagen, Denmark}



\date{\today}

\begin{abstract}
 We characterize parallel double quantum dot Josephson junctions based on closely-grown double nanowires bridged by in-situ deposited superconductors. 
 The parallel double dot behavior occurs despite the closeness of the nanowires and the potential risk of nanowire clamping during growth.
 By tuning the charge filling and lead couplings, we map out the simplest parallel double quantum dot Yu-Shiba-Rusinov phase diagram. 
Our quasi-independent two-wire hybrids show promise for the realization of exotic topological phases.

\end{abstract}

\maketitle


\section{\label{sec:level1}Introduction}

Double Rashba-nanowires bridged by superconductors are at the center of proposals for qubits\cite{Aasen2016Aug}, coupled subgap states\cite{Yao2014Dec} and exotic topological superconducting phases based on Majorana zero modes (MZMs)\cite{Beri2012Oct,Altland2013May,GaidamauskasPRL2014,Klinovaja2014Jul,EbisuProg2016,SchradePRB2017,ReegPRB2017,SchradePRL2018,ThakurathiPRB2018,DmytrukPRB2019,ThakurathiPRR2020,KotetesPRL2019,PapajPRB2019,HaimPhysRep2019}. Researchers have theorized on the existence of a topological Kondo phase in such wires when the bridging superconductor is in Coulomb blockade~\cite{Beri2012Oct,Altland2013May,GalpinPRB2014,PapajPRB2019} and, more recently, described a device hosting parafermions
~\cite{Klinovaja2014Jul}. 
Concretion of these proposals should benefit from material science developments resulting in improved nanowire/superconductor interfaces with low quasiparticle poisoning rates~\cite{Krogstrup2015Jan,Albrecht2016Mar,higginbotham2015parity}. 

These clean interfaces have been used in the investigation of MZMs in single nanowires~\cite{Deng2016Dec,Albrecht2016Mar} and, more recently, for coupling single and serial quantum dots (QDs) defined on single nanowires to superconductors to realize one and two-impurity Yu-Shiba-Rusinov (YSR) models
~\cite{EstradaSaldana2018Dec,EstradaSaldana2020Nov,Valentini2020Aug,EstradaSaldana2020Jul,Razmadze2020Sep}. 
YSR excitations, sometimes referred to as Andreev bound states~\cite{Buitelaar2002Dec,Deacon2010Feb,Deacon2010Mar,Pillet2010Dec,Lee2014Jan,Jellinggaard2016Aug,Li2017Jan,Lee2017May,Su2017Sep,Gramich2017Nov,grove2018yu,EstradaSaldana2020Nov,Su2020Jun,EstradaSaldana2020Jul,Valentini2020Aug,Prada2020Oct}, arise in the limit of large Coulomb charging energy, $U>\Delta$, where $\Delta$ is the size of the superconducting gap, as a result of the virtual excitation of a quasiparticle into the edge of the superconducting gap~\cite{Zitko2015Jan,Kirsanskas2015Dec}. This quasiparticle can exchange-fluctuate with a localized spin in the QD, and if the exchange coupling is strong (i.e., when the Kondo temperature, $T_\mathrm{K}$, is larger than $\sim 0.3\Delta$), the ground state (GS) transits from a doublet to a singlet~\cite{Satori1992Sep}. In Josephson junctions (JJs), this provokes a $\pi$-\textit{0} phase-shift change in the superconducting phase difference~\cite{Bauer2007Nov,Rozhkov2001Nov,van2006supercurrent,Cleuziou2006Oct,Grove-Rasmussen2007May,Jorgensen2007Aug,Eichler2009Apr,DeFranceschi2010Oct,Maurand2012Feb,Kim2013Feb,Delagrange2015Jun,Delagrange2018May,EstradaSaldana2018Dec,EstradaSaldana2020Nov,Meden2019Feb,EstradaSaldana2019Jul}. 

Devices which use pairs of QDs placed in a parallel configuration \cite{BabaAPL2015,Deacon2015Jul} and coupled to common superconducting leads have been extensively studied with the purpose of producing entangled electron states through Cooper pair splitting \cite{Hofstetter2009Oct,Das2012Nov,Baba2018Jun,Ueda2019Oct}. However, the behaviour of the switching current, $I_\mathrm{sw}$, in the presence of YSR screening \cite{grove2018yu,EstradaSaldana2020Nov,Kurtossypreprint2021} in parallel double QDs remains to be investigated. 

In this work, we characterize superconductivity in closely-grown pairs of InAs nanowires bridged by a thin epitaxial superconducting aluminum film deposited in-situ \cite{Kannepreprint2021}. To do so, we fabricate two side-by-side JJs out of one pair of nanowires and demonstrate that each nanowire hosts a single QD, through which supercurrent flows. 
From the charge stability diagram and magnetic field measurements, we establish that the interwire tunnelling is negligible with an upper bound of $\sim 10~\mu eV$. The YSR physics is analyzed through the gate dependence of the linear conductance and $I_\mathrm{sw}$, where we find that the common superconducting leads screen \textit{individually} each QD, hinting at individual YSR clouds instead of a single one extending over the two QDs. We furthermore show indications of supercurrent interference when the ground-state parity of the QDs is different, reminiscent of a superconducting quantum interference device (SQUID) at zero magnetic field.

The article is structured in sections. 
In Section II, we introduce the YSR double QD phase diagram and measurements of two double QD shells in different coupling regimes are presented establishing weak interdot coupling.
In Section III, we show signatures of interference between the supercurrents flowing through each junction. In Section IV, we demonstrate the YSR screening evolution of $I_\mathrm{sw}$. 
Finally, in Section V we present our conclusions and provide perspectives of our work.

\section{\label{sec1}Characterization of the parallel quantum-dot Josephson junction}

\begin{figure}[h!]
    \centering
    \includegraphics[width=1\linewidth]{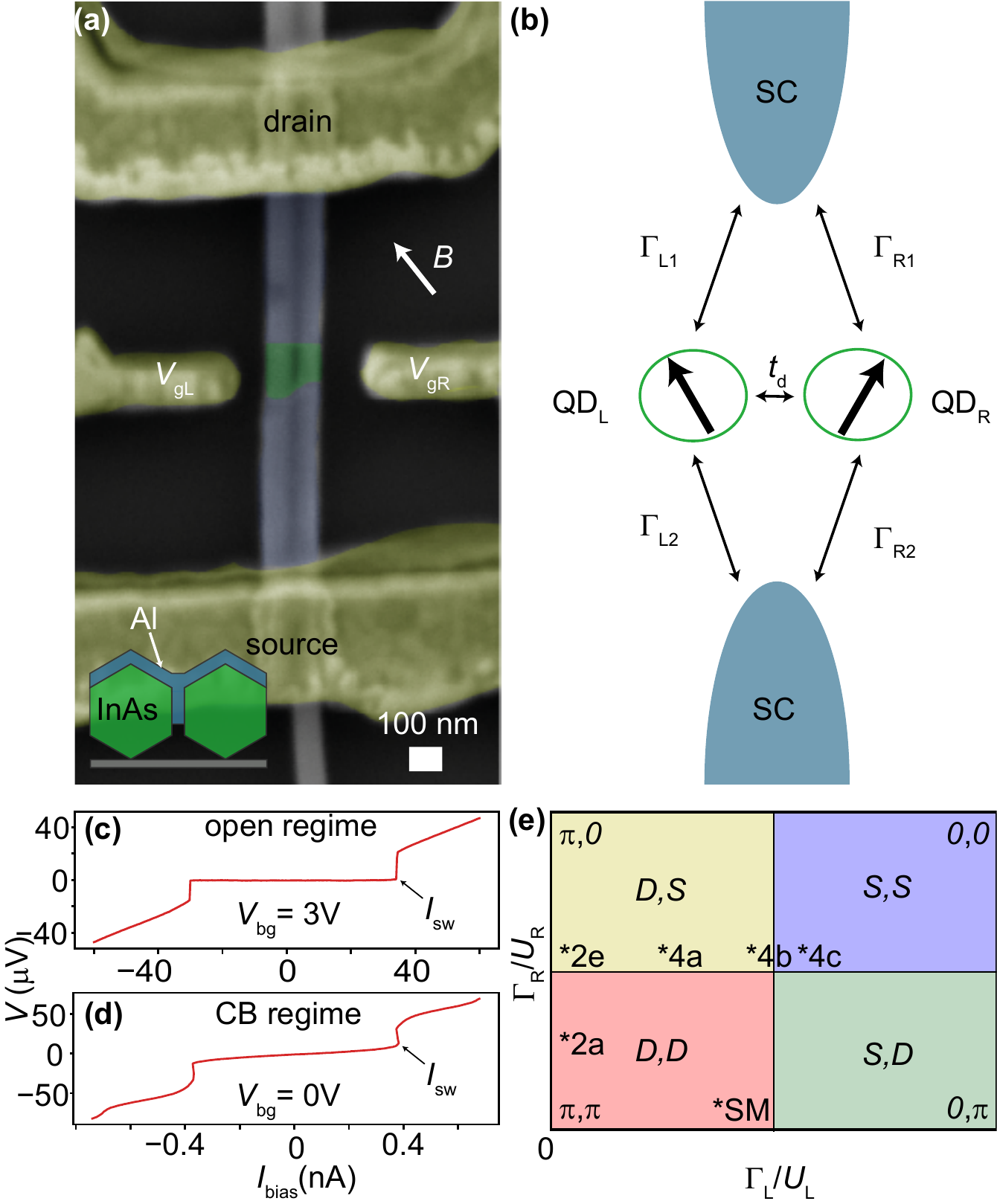}
    \caption{(a) Scanning electron micrograph of Device 1. Two nanowires with common superconducting leads form two parallel Josephson junctions. Side-by-side quantum dots serve as weak links for each JJ.
    The direction of an external in-plane magnetic field, $B$, when applied, is indicated by an arrow. In inset, a schematic cross-section of the double nanowire is shown, indicating facets of the nanowires covered by Al at the leads. (b) Sketch of the two side QDs coupled to two superconducting leads. Interdot tunnel coupling, $t_\mathrm{d}$, may be present. The GS parity of the left (L) and right (R) QDs is changed by tuning their level positions, $\epsilon_\mathrm{L}$ and $\epsilon_\mathrm{R}$, or by increasing the total tunnelling rates of each QD to the leads, $\Gamma_\mathrm{L}=\Gamma_\mathrm{{L1}}+\Gamma_\mathrm{{L2}}$ and $\Gamma_\mathrm{R}=\Gamma_\mathrm{{R1}}+\Gamma_\mathrm{{R2}}$. (c,d) $I_\mathrm{bias}-V$ curves measured at $V_\mathrm{bg}=\mathrm{3\, V}$ and $V_\mathrm{bg}=\mathrm{0\, V}$ showing switching current, $I_{\mathrm{sw}}$, in the open and in the Coulomb blockaded regimes, respectively. (e) Sketch of the GS phase diagram depending on the tunnelling rates $\Gamma_i$ ($i=$L,R) between the leads and the two QDs, when $t_\mathrm{d}=0$ and each QD has an unpaired electron. $D$ stands for doublet, and $S$ for singlet. The expected phase-shift in the Josephson current-phase relationship of each QD JJ, \textit{0} or $\pi$, is indicated. The qualitative $\Gamma_\mathrm{L}$, $\Gamma_\mathrm{R}$ positions of different shells from Figs.~\ref{fig2}, \ref{fig4} and SM, Section V (Device 2) are indicated by asterisks.}
    \label{fig1}
\end{figure}

In this section, we outline the device layout and demonstrate the Josephson effect and weak interdot tunnelling in Device 1. Data from an additional device (Device 2) is shown in Supplemental Material (SM).

Figure\,\ref{fig1}a shows a falsely-colored scanning electron microscopy (SEM) image of Device 1. 
Two 80-nm InAs nanowires (in green), grown close to each other in a molecular beam epitaxy chamber, and covered each on three of its facets by an in-situ deposited 17\,nm-thick layer of aluminum (in blue)\cite{Kannepreprint2021}, are individually picked with a micromanipulator and deposited on a Si/SiOx substrate. A resist mask is defined by electron beam lithography to selectively etch Al using the commercial etchant Transene-D, creating a parallel double JJ with $\approx$ 100\,nm-wide bare sections of the two nanowires as weak links. Ti/Au 5\,nm/250\,nm-thick contacts to Al and individual nanowire side-gates are deposited after a subsequent lithography step. Prior to the metal deposition step and without breaking vacuum, the Al native oxide is removed by argon milling to establish a good contact. The devices are measured in a dilution refrigerator at base temperature $T=\mathrm{ 30\,mK}$. 

QDs are formed when the two nanowires are brought near depletion with the use of the individual side-gate voltages, $V_{\mathrm{gL}}$ and $V_{\mathrm{gR}}$, and a global backgate voltage, $V_{\mathrm{bg}}$. The side gates are also used as plunger gates of the QDs. 
In Fig.\,\ref{fig1}b, we sketch the tunnelling rates of the QDs to the common superconducting leads (SC), $\Gamma_\mathrm{{L1}}$, $\Gamma_\mathrm{{L2}}$, $\Gamma_\mathrm{{R1}}$, and $\Gamma_\mathrm{{R2}}$, which can vary among different shells of the QDs and which can also be tuned by $V_{\mathrm{bg}}$. The QDs may also be coupled to each other by an interdot tunnel coupling, $t_\mathrm{d}$. We identify these different shells by the letters W, X, Y.
QD parameters extracted for these shells are given in Table \ref{tab1}. For an overview of the different shells explored, see SM, Section I.

The source and the drain contacts of the device each branch out into two leads as shown in Fig.~\ref{fig1}a, which we use to characterize the parallel JJs \cite{Guiducci2019Jun} in a four-terminal configuration by applying a current, $I_\mathrm{bias}$, from source to drain leads and measuring the voltage response, $V$, in a different pair of source and drain leads. In this way, we obtain $I_\mathrm{bias}-V$ curves which switch from a supercurrent branch at low $I_\mathrm{bias}$ to a high-slope dissipative branch at $I_{\mathrm{sw}}$. Two of such curves are shown in Fig.\,\ref{fig1}c,d for the open and Coulomb blockaded regimes, respectively. We measure $I_{\mathrm{sw}}$ up to 35\,nA in the former regime 
and up to approximately 500\,pA in the latter regime. In Coulomb blockade, the supercurrent branch shows a finite slope, $R_\mathrm{S}$, which increases with $\sim 1/I_\mathrm{sw}$ ; however, this does not affect our identification of $I_{\mathrm{sw}}$ as a jump in the curve down to 5 pA (see SM, Section II). In our analysis below (Sec.~III) we do not claim quantitative estimates of the critical current, $I_\mathrm{c}$ (which may be larger), but only address the qualitative behavior of $I_\mathrm{sw}$. 

As a guide to the different GS configurations accessed in this work, we show in Fig.\,\ref{fig1}e a sketch of the phase diagram of the parallel DQD JJ versus coupling to the leads when the two QDs have \textit{independent} GSs ($t_\mathrm{d}=0$).
The sketch corresponds to odd occupancy (1,1) of the QDs and it is valid for the large level-spacing regime, $\Delta E_i>U_i$, where i stands for left and right QDs. 
The independent-GS case is applicable to our device as most $I_\mathrm{sw}$ measurements are done away from the triple points of the QDs, where the effect of a finite $t_\mathrm{d}$ is negligible.
GS changes occur when the total tunnelling rates $\Gamma_\mathrm{{L,R}}$ of each of the QDs to the common superconducting leads surpass a threshold which depends on $U_\mathrm{L,R}/\Delta$~\cite{Lee2014Jan}, where $\Delta$ is the superconducting gap. Above this threshold, the spin of each QD is individually screened by the superconducting leads via the YSR mechanism~\cite{Zitko2011Feb,Yao2014Dec}.
For doublet GS, the current-phase relationship is $\pi$-shifted, e.g.~$I=I_\mathrm{c} \mathrm{sin}(\phi + \pi)$~\cite{Jorgensen2007Aug,van2006supercurrent,Maurand2012Feb,Cleuziou2006Oct} as indicated in Fig.~\ref{fig1}e.
\begin{table*}
\caption{\label{tab:table4}Parameters for shells W, X, Y of Device 1. 
The charging energies, $U_{\mathrm{L,R}}$, are extracted from Coulomb diamond spectroscopy. The total tunnelling rates of each QD, $\Gamma_{\mathrm{L,R}}$, are obtained by (a) fitting the even side of Coulomb diamonds in the normal state, or (b) from the full width at half maximum of the corresponding Coulomb peak. The Kondo temperature, $T_K$, is obtained by (c) fitting the Kondo peak (when applicable), or by (d) using the equation $T_K=\frac{1}{2k_B}\sqrt{\Gamma U} e^{\pi \epsilon_0 (\epsilon_0 + U)/\Gamma U}$, with $\Gamma_{\mathrm{L,R}}$, $U_{\mathrm{L,R}}$ as known values, and $\epsilon_0=\epsilon_\mathrm{L,R}$ the level position of the corresponding QD. Extraction methods are presented in detail in SM, Sec. III.}
\begin{ruledtabular}
\begin{tabular}{ccccccccccc}
 Shell&$U_\mathrm{L}$ $\mathrm{(meV)}$& $U_\mathrm{R}$ $\mathrm{(meV)}$ & $\Gamma_\mathrm{L}$ $\mathrm{(meV)}$ &$\Gamma_\mathrm{R}$ $\mathrm{(meV)}$ & $\frac{\Gamma_\mathrm{L}}{U_\mathrm{L}}$ &$\frac{\Gamma_\mathrm{R}}{U_\mathrm{R}}$ & $k_B T_{K_\mathrm{L}} \mathrm{(meV)}$ & $ k_B T_{K_\mathrm{R}}  \mathrm{(meV)}$  & $\frac{k_B T_{K_\mathrm{L}}}{0.3\Delta}$ & $\frac{k_B T_{K_\mathrm{R}}}{0.3\Delta}$\\ \hline
 W&$3.8\pm0.5$ &$2.3\pm 0.3$&$0.23\pm0.02^a$ & $0.6\pm 0.1^b$ & $0.06\pm0.01$ & $0.26 \pm0.05$ &$(3\pm0.3)\cdot 10^{{-5}^d}$&$0.03\pm0.01^d$  &$6\cdot 10^{-4}$ & $0.5$ \\
 X& $3.7\pm0.5$ &$1.1\pm0.3$ &$0.33\pm0.01^a$ &$0.55\pm0.1^d$ &$0.09\pm0.01$ & $0.5\pm0.1$ &$(8\pm1)\cdot 10^{{-5}^d}$ &$0.07-0.18^c$ &$0.001$ & $3.2$\\
 Y&$3.6\pm0.5$&$1.1\pm0.3$ & $1.05\pm0.01^a$  &$0.55\pm0.1^d$ &$0.3\pm0.04$  & $0.5\pm0.1$ & $0.06\pm0.02^d$ &$0.07-0.18^c$  &$1$ &$3.2$\\
\end{tabular}
\end{ruledtabular}
\label{tab1}
\end{table*}

\begin{figure}[h!]
    \centering
    \includegraphics[width=1\linewidth]{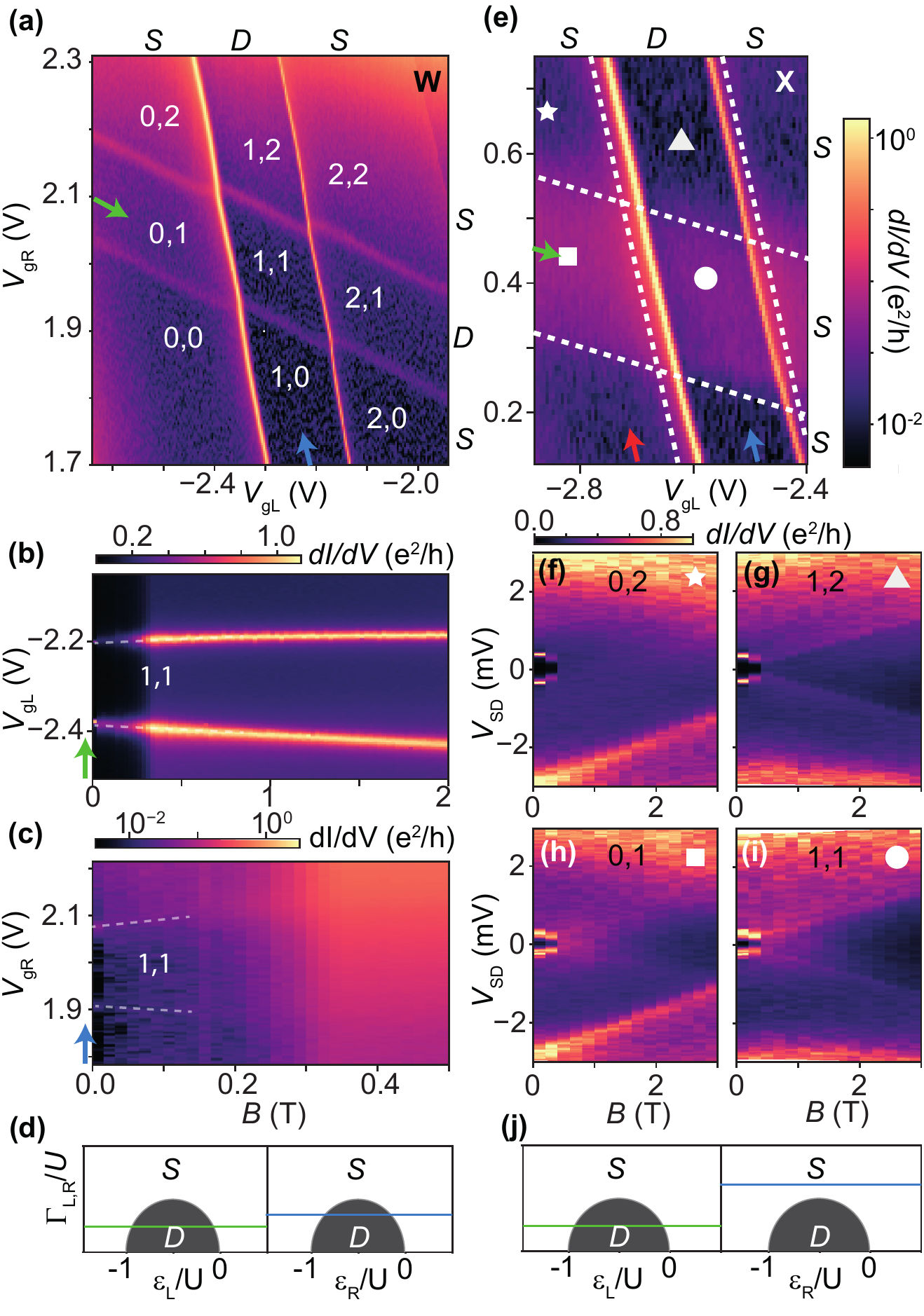}
    \caption{(a,e) Colormaps of two-terminal, voltage-biased zero-bias differential conductance, $dI/dV$, in the superconducting state for shells (a) W and (e) X vs.~left and right QD plunger gates. In (e), white dashed lines represent the position of the Coulomb lines measured at $B=2$ T.   
    (b,c) Zero-bias $dI/dV$ colormaps showing the magnetic field, $B$, dependence of parity transition lines which enclose the 1,1 charge sector in (a), vs.~plunger gate voltages of the (b) left and (c) right QDs, obtained by sweeping the gates along the green and blue arrows, shown in (a). For simplicity, only $V_\mathrm{gL}$ and $V_\mathrm{gR}$ are respectively shown.
    (f-i) Colormaps of $dI/dV$ vs.~magnetic field, $B$, and source-drain bias voltage, $V_{\mathrm{sd}}$, taken in four different charge sectors indicated by symbols in (e). (d,j) Pairs of phase-diagram sketches for independent left and right QDs. Horizontal color-coded lines in each pair indicate qualitatively $\Gamma_\mathrm{L} (\Gamma_\mathrm{R}$) vs.~left (right) QD level position $\epsilon_\mathrm{L} (\epsilon_\mathrm{R}$) in the stability diagrams of (a) and (e), respectively, following the arrows shown.}
    \label{fig2}
\end{figure}

To estimate $t_\mathrm{d}$, we first investigate via two-terminal voltage-biased differential conductance ($dI/dV$) measurements two shells corresponding to the two leftmost quadrants of the DQD phase diagram in Fig.~\ref{fig1}e. The two-terminal $dI/dV$ is recorded using standard lock-in amplifier techniques with an AC excitation of 2~$\mathrm{\mu V}$.
Figure \ref{fig2}a shows a colormap of $dI/dV$ at source-drain bias $V_\mathrm{SD}=0$ of shell W in the superconducting state versus $V_\mathrm{gL}$ and $V_\mathrm{gR}$, which represents the stability diagram of the two QDs in the weakly-coupled regime where $\Gamma_{\mathrm{L,R}} \ll U_{\mathrm{L,R}}$ (see Table \ref{tab1} for shell parameters). Since the slope of the supercurrent branch, $R_\mathrm{S}$, is empirically related in our device to $1/I_\mathrm{sw}$, we can use $R_\mathrm{S}=1/(dI/dV(V_\mathrm{SD}=0))$ as an indicator of $I_\mathrm{sw}$. This is particularly relevant in the Coulomb-blockade regime, when $I_\mathrm{sw}$ is small and $R_\mathrm{S}$ is significant (see SM, Section II). We observe approximately vertical and horizontal conductance lines which overlap and displace each other at their crossings, without exhibiting any significant bending. The displacement is a signature of a finite interdot charging energy, while the lack of bending indicates that $t_\mathrm{d} \approx 0$ (with an upper limit of 10 $\mu$V based on the width of the sharpest conductance lines). No signatures of CAR or of elastic co-tunnelling\cite{Scherubl2020Apr} 
are observed in this measurement. We interpret these lines as GS parity transition lines, which indicate changes of parity in the left and right QDs, respectively. The lines separate nine different and well-defined parity sectors. We assign corresponding effective left and right QD charges, $N_\mathrm{L},N_\mathrm{R}$, to each of these sectors based on the shell-filling pattern of the stability diagram in larger plunger-gate ranges (see SM, Section I). The charges obtained in this way are indicated in Fig.~\ref{fig2}a.

To assign GS parities to these nine sectors, and to determine independently if, in addition to interdot charging energy, there is 
a significant $t_\mathrm{d}$, 
we trace the evolution of the parity transition lines of the 1,1 charge sector against $B$. In the case of singlet GS; i.e., when the spins of the two QDs are exchange-coupled (finite $t_\mathrm{d}$), these lines are expected to come together with $B$. Instead, as shown in the zero-bias $dI/dV$ colormaps in Figs.~\ref{fig2}b,c, the parity transition lines enclosing the 1,1 charge sector split apart with $B$,
i.e., the two QDs are independent doublets, despite the relative closeness of the two nanowires. The splitting of the parity lines occurs both in the case when the parity of the left (right) QD is varied and the right (left) QD is kept in the doublet GS (see green and blue arrow, respectively, in Fig.~\ref{fig2}a).
The GS (singlet $S$ or doublet $D$) of the other eight charge sectors are indicated on the top and right exterior parts of the stability-diagram colormap in Fig.~\ref{fig2}a. 

Given the decoupling between the two 
QDs, we can approximate their phase diagrams by those of two independent single QDs. Neglecting the interdot charging energy, we sketch in Fig.~\ref{fig2}d the well-known single-QD phase diagrams for the GS of the left and right QDs versus QD level position, $\epsilon_\mathrm{L,R}$, and versus the total tunnelling rate of each QD to the leads, $\Gamma_{\mathrm{L,R}}$, over their charging energy, $U_{\mathrm{L,R}}$. 
The doublet dome has an upper height limit of $\Gamma_{\mathrm{L,R}}/U_{\mathrm{L,R}}=1/2$ in the infinite $\Delta$ limit, and its height decreases in the $U \gg \Delta$ limit (i.e., the YSR regime) to which our QDs belong~\cite{Meng2009Jun,Lee2017May}. 
In the left phase diagram, the horizontal green line which crosses the doublet dome indicates a cut where $\epsilon_\mathrm{L}$ is varied and $\epsilon_\mathrm{R}$ is kept fixed such that the GS parity of the right QD is a doublet, and the GS parity of the left QD is variable. This line represents schematically the gate trajectory in Fig.~\ref{fig2}b, as indicated with the green arrow, which is collinear to the green arrow in Fig.~\ref{fig2}a, and which varies the parity of the left QD as $S$-$D$-$S$ while keeping the parity of the right QD as $D$. A similar relation exists between the horizontal blue line in the right phase diagram, and the gate trajectory (blue arrow) in Fig.~\ref{fig2}c, also collinear to the corresponding arrow in Fig.~\ref{fig2}a. From these phase diagrams, we note that parity transitions are strictly equal to Coulomb degeneracies only at zero $\Gamma_{\mathrm{L,R}}$). The measurements above confirm the expected DQD behavior for low lead couplings, which shows a $D$,$D$ ground for charge state 1,1 corresponding to the lower left quadrant of the phase diagram in Fig.~\ref{fig1}e.

Next, we investigate a differently coupled shell (shell X) which belongs to the upper left quadrant of phase diagram in Fig.\ \ref{fig1}e. Figure \ref{fig2}e shows the zero-bias $dI/dV$ colormap in the superconducting state vs.~the plunger gates of the two QDs of shell X.
The two horizontal GS-parity transition lines, which bounded the green trajectory in the case of shell W, are absent in the case of shell X, and are instead replaced by a band of enhanced conductance. The conductance band is cut two times by approximately vertical conductance lines, which correspond 
to GS-parity transition lines of the left QD.

The parity of the band of enhanced conductance in the stability diagram is determined from the $B$-evolution of the differential conductance in the normal state versus $\mathrm{V_{sd}}$ at two fixed gate voltages. 
These two gate voltages are indicated by a square (charge states 0,1) and a circle (1,1) in the stability diagram, and their $B$ dependence is respectively shown in Figs.~\ref{fig2}h,i. As a control experiment, the $B$ dependence for two fixed gate voltages above the conductance band 
indicated by a star (0,2) and a triangle (1,2) in the stability diagram, is shown in Figs.~\ref{fig2}f,g. The four measurements show closing of the superconducting gap at $B=0.4$ T, which is consistent with the jump in the zero-bias $dI/dV$ signal in Figs.~\ref{fig2}b,c at $B \approx 0.4$ T. However, in the normal state, whereas Figs.~\ref{fig2}g-i (1,2 0,1 1,1) display conductance steps near zero-bias which split with $B$ field, in Fig.~\ref{fig2}f there is no such splitting, consistent with even filling of both dots. 
We assign effective QD charge numbers to the charge stability diagram from a $B=2$ T measurement (see SM, Section III) and overlay the Coulomb lines obtained, which delimit the nine charge sectors (white dashed lines in Fig.~\ref{fig2}e).

We note an additional important difference in the data of the low-bias splitting states. Whereas in Fig.~\ref{fig2}g (1,2) the splitting can be traced back to zero bias at $B=0$, in Fig.~\ref{fig2}h (0,1) the splitting is traced to zero bias only at a finite field of $\approx 1$ T. 
The pair of features whose splitting can be traced to a $B=0$ onset in Fig.~\ref{fig2}g (1,2) correspond to co-tunnelling steps of the 
odd-occupied left QD experiencing Zeeman splitting. In turn, the pair of features which starts to split at 1 T in Fig.~\ref{fig2}h corresponds to the Zeeman splitting of a Kondo resonance in the right QD. The splitting ensues when $E_\mathrm{Z} \sim k_B T_\mathrm{K}^R$.\cite{Kogan2004Oct} Notice that the Kondo resonance is also visible in the data after the gap closure at $B=0.4~\mathrm{T}$. From the splitting, we find a $g$-factor $g\sim \mathrm{8.5\pm 0.1}$. Table~\ref{tab1} shows that $k_BT_{K_\mathrm{R}}>0.3\Delta$ for shell X, which is consistent with a YSR singlet state in the right QD in the superconducting state.

The $B$-dependence data in Figs.~\ref{fig2}f-i 
therefore 
allows us to assign the GS to the QDs, $D$ or $S$, in each of the nine sectors in Fig.~\ref{fig2}e. We indicate schematically by a green and blue horizontal line in the two individual-QD phase-diagram cartoons in Fig.~\ref{fig2}j the GS along the gate trajectories collinear to the same-colored arrows in the colormap of Fig.~\ref{fig2}e.
The green (blue) gate trajectory, which goes along (perpendicular to) the band of enhanced conductance intersects twice (goes above) the doublet dome, leading to two (zero) parity transitions.

\section{\label{sec2}Supercurrent interference for different quantum-dot parities}

\begin{figure}[t!]
    \centering
    \includegraphics[width=0.85\linewidth]{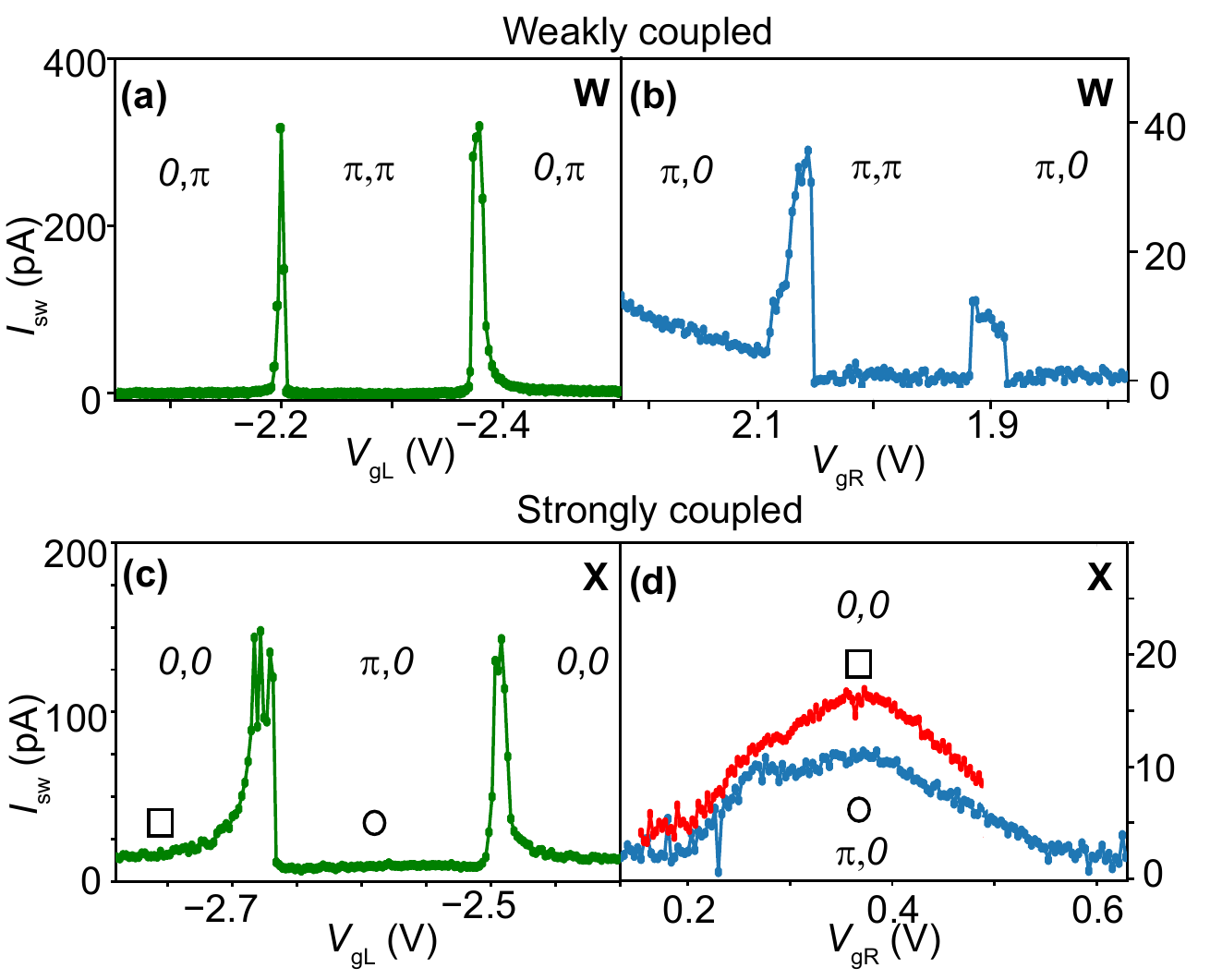}
    \caption{(a-d) Extracted 
    $I_{\mathrm{sw}}$ vs.~plunger-gate voltage trajectories collinear to same-colored arrows in (a,b) Fig.~\ref{fig2}a, shell W, and (c,d) Fig.~\ref{fig2}e, shell X. In (d), two traces are shown to illustrate the decrease in $I_{\mathrm{sw}}$ as a consequence of the subtracting effect of a $\pi$ phase-shift in one of the QD Josephson junctions. The red curve is offset on the gate axis in order to correct for the cross-talk between the gates and the QDs.}
    \label{fig3}
\end{figure}

We switch back to the four-terminal measurement configuration to correlate the intrinsic phase of each JJ with the magnitude of $I_{\mathrm{sw}}$.
In Fig.~\ref{fig3}, we show $I_{\mathrm{sw}}$ versus plunger gate voltages, where $I_{\mathrm{sw}}$ is extracted in a similar fashion as in Fig.~\ref{fig1}d.
In Figs.~\ref{fig3}a,c (\ref{fig3}b,d), the plunger gate voltages are swept along trajectories which vary the occupation in the left (right) QD while keeping the occupation of the right (left) QD fixed, following the green (red, blue) arrows in Figs.~\ref{fig2}a,e, i.e. for shell W and X, respectively. 
For reference, we assign the expected phase-shift in the current-phase relationship, $\pi$ or $\textit{0}$, based on the measured GS parities of the two QDs. This phase-shift is accurate when at least one QD is in Coulomb blockade. The value of $I_{\mathrm{sw}}$ at the parity transitions may include contribution due to presence of bound states crossing zero-energy. Hence, the magnitude of $I_{\mathrm{sw}}$ on transitions should not be taken into account.

\begin{figure}[t!]
    \centering
    \includegraphics[width=0.85\linewidth]{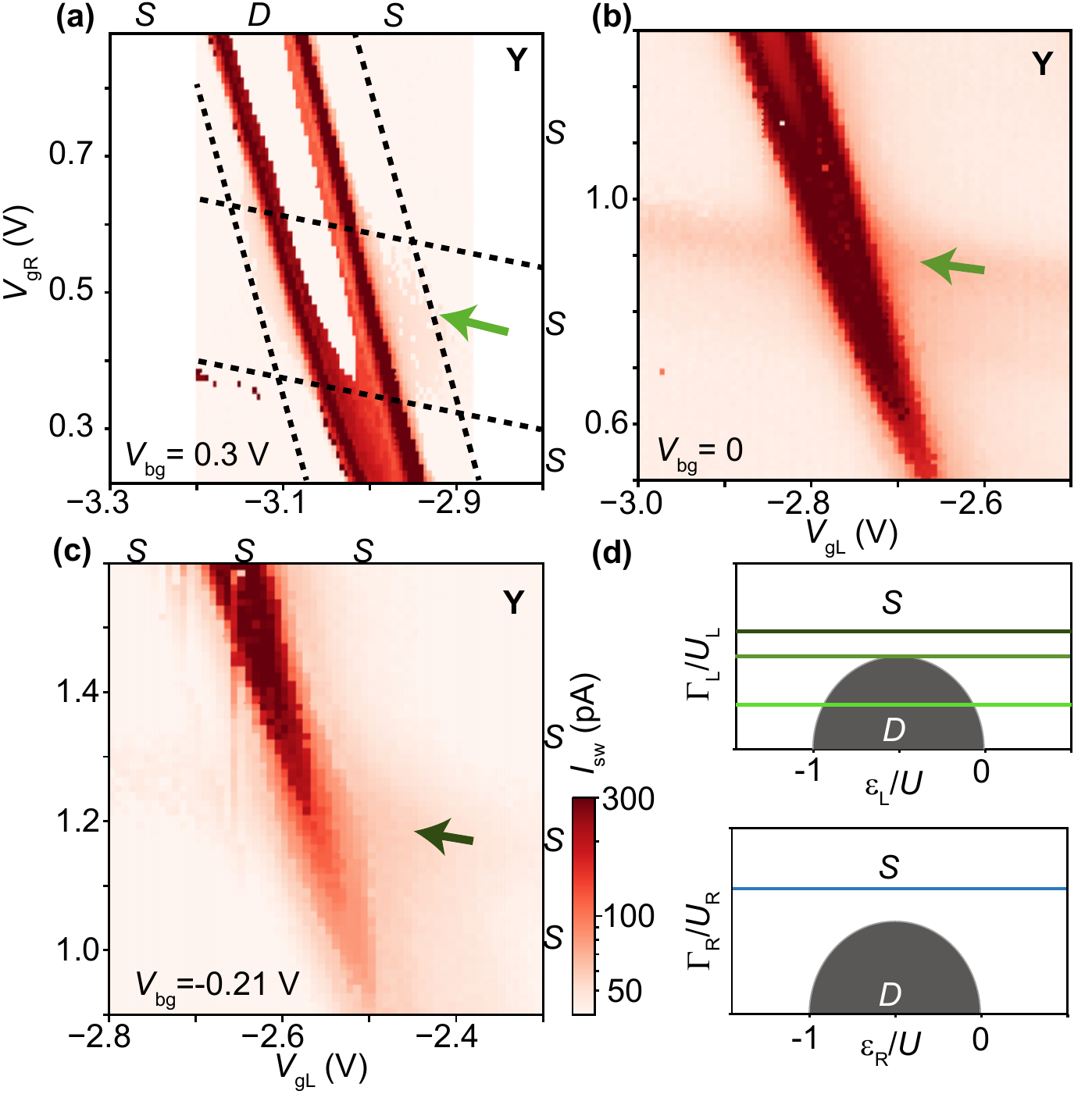}
    \caption{(a-c) Colormaps of 
    $I_{\mathrm{sw}}$ as a function of the plunger gates of the two QDs, taken at three different $V_{\mathrm{bg}}$ values in shell Y. In (a), Coulomb lines positions (black dashed lines) are obtained from a normal-state two-terminal differential conductance measurement at $B=2$ T. To keep shell Y in frame, the effect of $V_{\mathrm{bg}}$ has been compensated by changing $V_{\mathrm{gL}}$ and $V_{\mathrm{gR}}$. In (a) and (c), the GS of the two independent QDs is indicated on the exterior side of the colormaps. (d) Independent-QD phase-diagram cartoons as function tunnelling rate $\Gamma_{\mathrm{L,R}}$ and QD level position, $\epsilon_{\mathrm{L,R}}$, for the left QD (top panel) and right QD (lower panel). In the top panel, green-shaded horizontal lines indicate qualitatively $\Gamma_\mathrm{L}$ in directions collinear to the arrows of the same color in (a-c). The blue line indicates qualitatively $\Gamma_\mathrm{R}$ in (a-c).}
    \label{fig4}
\end{figure}

The common phenomenology in the data is as follows. After a smooth build-up of $I_{\mathrm{sw}}$ towards a $\textit{0} \to \pi$ transition, the current abruptly drops right at the edge of the $\pi$ domain, resulting in an asymmetric $I_{\mathrm{sw}}$ peak~\cite{Jorgensen2007Aug}. A pair of asymmetric peaks is seen in the data in Figs.~\ref{fig3}a-c, as one of the QDs experiences parity transitions and therefore a sequence of $\textit{0}-\pi-\textit{0}$ phase-shift changes. If the parity stays unchanged, such peaks are absent, as in Fig.~\ref{fig3}d. Instead, $I_{\mathrm{sw}}$ is smoothly enhanced towards odd occupation of the right QD, which is YSR-screened (i.e., $k_BT_K > 0.3\Delta$)~\cite{Maurand2012Feb}.
Interestingly, when comparing the red and blue traces in Fig.~\ref{fig3}d, which correspond to different phase shift ($\pi$ and \textit{0}, respectively) in the JJ formed by the left QD, we observe that $I_{\mathrm{sw}}$ is stronger near $V_{\mathrm{gR}}=0.4$ V. Note that $V_{\mathrm{gR}}=0.4$ V corresponds to the 1,1 charge state for the blue trace, and to the 0,1 charge state for the red trace. The exact magnitude of $I_{\mathrm{sw}}$ in that gate value for the red and blue curve is consistent with what is found in Fig.~\ref{fig3}c in the ($\square$) and ($\bigcirc$) respectively.   
We can interpret the reduction in $I_{\mathrm{sw}}$ at $V_{\mathrm{gR}}=0.4$ V in the blue trace with respect to the red trace by considering the double nanowire device as a SQUID at zero threaded magnetic flux~\cite{Cleuziou2006Oct,van2006supercurrent,Maurand2012Feb}. The $I_{\mathrm{c}}$ of a SQUID with a sinusoidal current-phase relation can be written as~\cite{Maurand2012Feb}

\begin{eqnarray}
I_\mathrm{c}=\sqrt{(I_\mathrm{{c1}}-I_\mathrm{{c2}})^2 +4I_\mathrm{{c1}}I_\mathrm{{c2}}\left\lvert cos\Big(\pi\frac{\phi}{\phi_0}+\frac{\delta_1+\delta_2}{2}\Big) \right\rvert^2}
\end{eqnarray}

where $I_{\mathrm{c1,c2}}$ are the critical currents of the two JJs, $\phi$ is the threaded magnetic flux and $\delta_{1,2}$ are the intrinsic phase shifts (\textit{0} or $\pi$) of the junctions. As a result, for $\phi=0$ the total $I_\mathrm{c}$ is given by $I_{\mathrm{c}\square}= I_{\mathrm{c1}}+I_{\mathrm{c2}} $ when the DQD is in the \textit{0,0} phase, and $I_{\mathrm{c}\bigcirc}= I_{\mathrm{c1}}-I_{\mathrm{c2}}$ in the $\pi$,\textit{0} phase. These equations may explain the findings in Fig.\,\ref{fig3}c,d, as $I_{\mathrm{sw}}$ is enhanced when both JJs have the same intrinsic phase, and it is weaker when the two JJs have different phase.

\section{\label{sec3}Screening evolution of switching current}
Finally, we demonstrate individual control of the couplings between the SC leads and the QDs, realizing the transition from the upper left (one screened spin in 1,1) to upper right quadrant (both spins screened) in the YSR phase diagram depicted in Fig.~\ref{fig1}e. Whereas the changes in GS parity in Fig.~\ref{fig2} occurred primarily by changing the side-gate voltages to go from shell W to shell X, here the changes occur within a unique shell. This is done in a shell identified as Y, using $V_{\mathrm{bg}}$ as a tuning knob of $\Gamma_\mathrm{L,R}$. In Figs.~\ref{fig4}a-c, we show colormaps representing parity stability diagrams at different $V_{\mathrm{bg}}$ analogous to those in Figs.~\ref{fig2}a,e; however, instead of plotting a measurement of voltage-biased $dI/dV$, we directly plot a four-terminal measurement of $I_{\mathrm{sw}}$ vs.~plunger-gate voltages. To obtain each colormap, we measure the $I_\mathrm{bias}-V$ characteristic at each plunger gate voltage coordinate (i.e., at each pixel in the colormap) and extract $I_{\mathrm{sw}}$ as in the example in Fig.~\ref{fig1}d. 

In Fig.~\ref{fig4}a, the $I_{\mathrm{sw}}$ parity stability diagram shows two $I_{\mathrm{sw}}$ peaks which correspond to two parity transitions of the left QD. The lack of right-QD parity transition lines indicates that the right QD is YSR-screened. We corroborate that this is indeed the case from a measurement of $T_{K_\mathrm{R}}$ at $B=0.4$ T in the normal state, and we find $k_BT_{K_\mathrm{R}}>0.3\Delta$ (see Table \ref{tab1}). We also note that, although faintly-visible here, a two-terminal $dI/dV$ measurement of the stability diagram in otherwise the same conditions as here displays an horizontal band of (weakly) enhanced conductance, which is the same phenomenology identified in Fig.~\ref{fig2}d with YSR spin-screening. However, the enhancement is weak enough to preclude resolution of $I_{\mathrm{sw}}$, and therefore a similar band of $I_{\mathrm{sw}}$ does only show at the right part of Fig.~\ref{fig4}a ($\mathrm{V_{gL}}\approx-2.95$ V, $\mathrm{V_{gR}}\approx0.45$ V). 

Reducing $\mathrm{V_{bg}}$ alters the $I_{\mathrm{sw}}$ parity stability diagram by bringing the two $I_{\mathrm{sw}}$ peaks (parity lines) of the left QD closer together, as shown in Fig.~\ref{fig4}b. Note that a faint, approximately horizontal band of $I_{\mathrm{sw}}$ is observed along the direction pointed by the dark-green arrow, which comes as a result of enhancement of $I_{\mathrm{sw}}$ due to YSR spin-screening of the right QD. In Fig.~\ref{fig4}c, further reduction of $\mathrm{V_{bg}}$ leads to merging of the parity lines into a vertical band of $I_{\mathrm{sw}}$ across the whole plot. At this point, the spins of both QDs are YSR-screened into singlets. We have therefore traced the phase diagram shown in Fig.~\ref{fig1}e, where either one spin of a QD or both are screened by the YSR mechanism, triggering a phase change in the current-phase relation of the JJs. Additional data on the magnetic field dependence of this shell can be found in the SM, section IV.

\section{\label{sec6}Conclusions \& Outlook}
In conclusion, we have demonstrated parallel quantum-dot Josephson junctions fabricated out of a double-nanowire platform in which the nanowires are bridged by an in-situ deposited superconductor. We mapped out the parallel quantum dot YSR phase diagram via conductance and switching current measurements showing the tunability of the ground state of each JJ from doublet to singlet. 
The analysis also revealed that the nanowires are predominantly decoupled with an upper bound on the interwire/dot tunnel coupling in the order of $t_\mathrm{d} \le 10~\mu$eV for the specific charge states studied in two devices. Finally, we showed indications of switching current addition and subtraction via appropriate choice of ground states of the two dots involving the YSR singlet state, i.e., $0,0$ and $\pi,0$ (phase difference) regimes, respectively. 

The above observations of basic superconducting properties in in-situ made hybrid double nanowire material open up for more advanced experiments addressing a number of recent theoretical proposals. In parallel double-quantum-dot Cooper-pair splitters~\cite{Hofstetter2009Oct,Herrmann2010Jan}, the CAR mechanism responsible for the splitting is weakened by an increase in the distance between the tunnelling points from the superconductor into the two quantum dots~\cite{Recher2001Apr}. The closeness of the nanowires set by growth\cite{Kannepreprint2021} and the cleanness of the Al-InAs interface may turn out beneficial for CAR, which is also the basis for creating coupled YSR states in these systems \cite{Yao2014Dec,Scherubl2019}. The latter is investigated in a parallel work on the same hybrid double nanowire material\cite{Kannepreprint2021} as used in this work\cite{Kurtossypreprint2021}. 
The hybrid double nanowires are furthermore interesting for realizing several species of topological subgap states\cite{GaidamauskasPRL2014,Klinovaja2014Jul}. 
For finite CAR, the requirements for entering the topological regime hosting Majorana bound states have been shown to be lowered \cite{SchradePRB2017,DmytrukPRB2019}, and 
parafermions may be achieved in a regime where CAR dominates over local Andreev processes\cite{Klinovaja2014Jul}. In superconducting islands fabricated in our hybrid double nanowires, the topological Kondo effect may be pursued
\cite{Beri2012Oct,Altland2013May,PapajPRB2019}. In double-nanowire Josephson junctions as here demonstrated, however, in the topological regime, non-standard types of Andreev bound states have been predicted \cite{KotetesPRL2019}. As an ending remark, we note that double nanowires can also be made with a full superconducting shell \cite{Kannepreprint2021,VekrispreprintLP2021}, relevant for investigating flux-induced subgap states\cite{Valentini2020Aug,Vaitiekenas2020Mar,PenarandaPhysRevRes2020}.

\section{\label{sec7}Acknowledgements}
We thank Gorm Steffensen, Jens Paaske, Michele Burrello and Constantin Schrade for useful discussions. We thank Xinyan Wang for experimental contribution. We acknowledge the support of the Sino-Danish Center, the European Union’s Horizon 2020 research and innovation program under the Marie Sklodowska-Curie Grant Agreement No. 832645, QuantERA “SuperTop” (NN 127900), the European Union’s Horizon 2020
research and innovation programme FETOpen Grant No. 828948 (AndQC), the Danish National Research Foundation, Villum Foundation Project No. 25310, the Ministry of Science and Technology of China through the National Key Research and Development Program of China (Grant Nos. 2017YFA0303304, 2016YFA0300601), the National Natural Science Foundation of China (Grant No. 11874071), the Beijing Academy of Quantum Information Sciences (Grant No. Y18G22), Carlsberg Foundation and the Independent Research Fund Denmark.

\bibliography{DNW_JJ}
\end{document}